ASTRONOMY

# Late formation of silicon carbide in type II supernovae

Nan Liu,* Larry R. Nittler, Conel M. O'D. Alexander, Jianhua Wang

We have found that individual presolar silicon carbide (SiC) dust grains from supernovae show a positive correlation between $^{49}$Ti and $^{28}$Si excesses, which is attributed to the radioactive decay of the short-lived ($t_{1/2}$ = 330 days) $^{49}$V to $^{49}$Ti in the inner highly $^{28}$Si-rich Si/S zone. The $^{49}$V-$^{49}$Ti chronometer shows that these supernova SiC dust grains formed at least 2 years after their parent stars exploded. This result supports recent dust condensation calculations that predict a delayed formation of carbonaceous and SiC grains in supernovae. The astronomical observation of continuous buildup of dust in supernovae over several years can, therefore, be interpreted as a growing addition of C-rich dust to the dust reservoir in supernovae.

## INTRODUCTION

In the last decade, our understanding of dust formation in the universe has been revolutionized by the discovery of massive amounts of dust in young galaxies (1–5). The young ages of these galaxies (1 to 100 million years) overwhelmingly suggest rapidly evolving type II supernovae (SNe) as efficient dust producers (for example, 5). This possibility is further supported by theoretical predictions that 0.08 to 1 solar masses ($M_⊙$) of dust can condense within a few years after SN explosions (6–11). Earlier observations of the amount of dust formed in SNe were obscured by the presence of preexisting interstellar dust in front of the remnants (12, 13). Recent detections based on Herschel and ALMA (Atacama Large Millimeter/submillimeter Array) observations with improved spatial resolutions have confirmed that 0.1 to 0.5 $M_⊙$ of dust formed within several SNe (14–16), but lower amounts of dust have been observed in other recent SNe shortly after the explosions and, therefore, cannot account for the dust contents in young galaxies (17–19). These discrepant observations could be reconciled if dust is continuously built up over several years (20, 21). A test of this idea requires long-term observations of in situ dust formation starting immediately after SN explosions, which to date are available for only a few SNe (20). Here, we use an alternative method, laboratory isotopic measurements of presolar SN dust grains extracted from meteorites, to infer the timing of dust production in SNe.

Various types of dust grains that condensed in presolar evolved stars and SNe have been found in primitive meteorites, including graphite, silicon carbide (SiC), nanodiamond, silicon nitride, oxides, and silicates (22). Recovered dust grains with isotopic signatures indicating an SN origin are dust that formed in SN ejecta and survived destructive processes in the SNe, interstellar medium, and protosolar nebula before their incorporation into the host meteorites that formed early in solar system history. The so-called X SiC grains (1 to 2% of all presolar SiC grains) have been extensively studied and are characterized by large $^{28}$Si excesses and the initial presence of short-lived $^{44}$Ti with a half-life of 60 years, both of which point to an SN origin (23, 24). Nucleosynthetic and thermodynamic equilibrium calculations show that X grains must have incorporated materials from at least the inner Si/S zone and the outer C-rich He/C zone (also see the Supplementary Text) (25, 26) to explain their large $^{28}$Si and $^{44}$Ti excesses (Fig. 1) and to condense in a C-rich environment, respectively (27). It is also noteworthy to point out the following two aspects: (i) Selective mixing of materials from the inner Si/S zone and the outer C-rich zones, as suggested by the

Department of Terrestrial Magnetism, Carnegie Institution for Science, Washington, DC 20015, USA.
*Corresponding author. Email: nliu@carnegiescience.edu

isotopic signatures of presolar X grains, may be supported by the observation of Rayleigh-Taylor instabilities in three-dimensional simulations of core-collapse SNe [for example, (28)]. (ii) Electrons arising from $^{56}$Co decay could dissociate the stable CO molecule, allowing carbonaceous grains to grow even in O-rich conditions [for example, (29)]. However, quasi-equilibrium calculations (30) show that, although graphite could be stable in O-dominated SN zones, SiC is not stable under such O-rich conditions. On the other hand, the short-lived ($t_{1/2}$ = 330 day) $^{49}$V, which decays to $^{49}$Ti, is abundantly made in the inner Si/S zone, along with α-nuclides such as $^{28}$Si and $^{48}$Ti (Fig. 1). Vanadium and Ti also readily condense into SiC, so X grains could have incorporated large amounts of either live $^{49}$V or its daughter product $^{49}$Ti depending on their time of formation. Thus, the $^{49}$V-$^{49}$Ti systematics of X grains provides a potential chronometer for grain formation in SNe.

Previous attempts to use this chronometer, however, produced controversial and conflicting results. The first systematic study of Ti-V isotopes in X grains observed a good correlation between their $δ^{49}$Ti/$^{48}$Ti (see Table 1 for the definition of δ notation) and $^{51}$V/$^{48}$Ti values (hereafter referred to as Ti-V correlation), suggesting that condensation of X grains occurred within several months of SN explosions (31). The inferred early formation of X grains contradicts recent model predictions of late SiC formation in SNe (32), but a subsequent study did not reproduce the Ti-V correlation (Fig. 2A), calling the inferred early formation timing into question (33). In turn, the lack of a Ti-V correlation would imply either grain formation after the decay of most $^{49}$V or sources of $^{49}$Ti other than the decay of $^{49}$V (33). There is another non-negligible source of $^{49}$Ti for X grains: the outer C-rich He/C zone, where $^{49}$Ti is abundantly made by a neutron-capture process (Fig. 1). The $δ^{29}$Si/$^{28}$Si and $δ^{30}$Si/$^{28}$Si values of the Si/S zone are nearly −1000 per mil (‰) because the α-nuclide $^{28}$Si is abundantly made by an α-capture process within this zone. X grains, however, show a wide range of Si isotope ratios (for example, fig. S1), reflecting variable contributions to their Si budgets from the less $^{28}$Si-rich He/C zone. The $^{49}$Ti budgets of X grains could also be affected by the variable contributions of the He/C zone. Using Ti isotopes to constrain the formation timing of X grains in SNe, therefore, has been hampered by two difficulties: (i) It is difficult to accurately estimate the mixing ratios for different X grains as a result of the large uncertainties in SN nucleosynthetic predictions for the He/C zone (25, 31, 33), and (ii) potential Ti-V elemental fractionation could occur within the Si/S or He/C zones before zonal mixing, for example, fractionation of S relative to Si by preferential capture of SiS molecules from the Si/S zone into growing SiC grains was previously suggested to explain the large $^{32}$S excesses found in some X grains (34).

We use a different approach here to circumvent these problems and derive a more robust estimate of the timing of SN SiC grain formation. We have obtained Ti-V isotope ratios in 16 large X grains with a range of Si isotope ratios using a NanoSIMS 50L ion microprobe (Table 1). The new grain data are consistent with the few available literature data (23, 24, 31, 33, 35, 36) that have comparable precisions (Fig. 2B). All the grain data together define a negative correlation ($R = -0.58$) between their $\delta^{49}$Ti and $\delta^{30}$Si values (see the Supplementary Text). The correlation clearly links the $^{49}$Ti excesses of the grains to contributions from the Si/S zone in which the Si isotopes are essentially pure $^{28}$Si. That is, the negative correlation means that the $^{49}$Ti budgets of X grains are predominantly controlled by the decay of $^{49}$V that was produced in the Si/S zone. The lack of any observed correlation between $^{51}$V/$^{48}$Ti and $\delta^{30}$Si for the same set of X grains (Fig. 2C) clearly demonstrates that the large range of $^{51}$V/$^{48}$Ti in the X grains was not caused by variable mixing ratios between the Si/S and He/C zones. Note that X grains from the Qingzhen enstatite meteorite in the study of Lin et al. (33) show distinctively low $^{51}$V/$^{48}$Ti ratios relative to those from the Murchison carbonaceous meteorite, which could indicate preferential removal of V from the Qingzhen X grains by early solar nebular or parent-body processes, or two groups of SN grains from different SN ejecta or from the same SN ejecta but at different times.

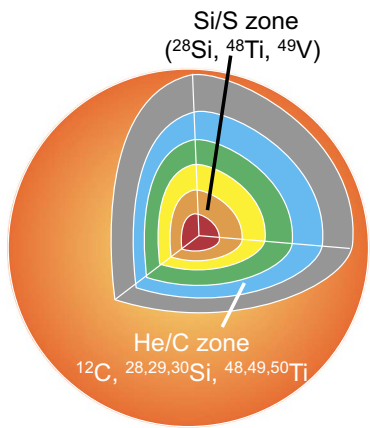

**Fig. 1. Schematic diagram of the "onion-shell" internal structure of a pre-SN massive star.** Zones are labeled by their most abundant elements (26). Neutron capture taking place in the outer C-rich He/C zone converts $^{28}$Si and $^{48}$Ti to neutron-rich Si and Ti isotopes, respectively, whereas α-capture in the inner Si/S zone overproduces α-nuclides, including $^{28}$Si and $^{48}$Ti. Abundant short-lived $^{49}$V is also made in the Si/S zone.

**Table 1. Carbon, N, Si, Al, and Ti isotopic compositions of type X SiC grains in Murchison.** Uncertainties are 1σ. A, agglomerate-like morphology; S, single grains. Four grains had too low Ti concentrations so that their Ti isotope ratios could not be determined.

| Grain | Size (μm) | Morphology | $^{12}$C/$^{13}$C | $^{14}$N/$^{15}$N | $\delta^{29}$Si* (‰) | $\delta^{30}$Si (‰) | $^{26}$Al/$^{27}$Al (×1000) | $\delta^{46}$Ti (‰) | $\delta^{47}$Ti (‰) | $\delta^{49}$Ti (‰) | $\delta^{50}$Ti (‰) | V/Ti |
|---|---|---|---|---|---|---|---|---|---|---|---|---|
| M1-A3-G320 | 1.6 × 2.8 | A | 224 ± 2 | 51 ± 1 | −273 ± 5 | −448 ± 8 | 392 ± 3 | 39 ± 64 | 113 ± 93 | 400 ± 19 | 228 ± 10 | 0.311 |
| M1-A7-G961 | 1.1 × 1.0 | A | 164 ± 2 | 113 ± 4 | −399 ± 5 | −544 ± 8 | 268 ± 1 | 40 ± 47 | −6 ± 70 | 350 ± 18 | −8 ± 8 | 0.094 |
| M1-A8-G137 | 1.9 × 1.3 | A | 149 ± 2 | 65 ± 2 | −286 ± 6 | −445 ± 17 | 350 ± 1 | −17 ± 72 | −84 ± 86 | 402 ± 19 | 206 ± 10 | 0.116 |
| M1-A9-G921 | 1.3 × 1.2 | S | 211 ± 3 | 54 ± 1 | −316 ± 8 | −464 ± 12 | 230 ± 1 | −4 ± 89 | −55 ± 103 | 528 ± 21 | 220 ± 12 | 0.054 |
| M2-A1-G421 | 0.6 × 0.6 | S | 145 ± 4 | 126 ± 3 | −237 ± 7 | −375 ± 15 | 153 ± 2 | 45 ± 29 | 0 ± 32 | 508 ± 21 | 191 ± 11 | 0.273 |
| M2-A1-G674 | 2.5 × 2.6 | A | 120 ± 3 | 87 ± 1 | −210 ± 6 | −290 ± 8 | 303 ± 1 | −8 ± 17 | −23 ± 22 | 304 ± 17 | 198 ± 10 | 0.139 |
| M2-A1-G904 | 2.6 × 2.7 | S | 100 ± 3 | 70 ± 1 | −168 ± 6 | −341 ± 12 | 277 ± 1 | 15 ± 37 | 13 ± 22 | 261 ± 29 | 186 ± 29 | 0.047 |
| M2-A1-G974 | 2.1 × 2.5 | A | 139 ± 4 | 116 ± 2 | −222 ± 6 | −343 ± 10 | 203 ± 1 | −2 ± 27 | −6 ± 12 | 158 ± 15 | 188 ± 10 | 0.017 |
| M2-A2-G373 | 0.9 × 0.9 | S | 149 ± 6 | | −183 ± 17 | −186 ± 25 | 39 ± 1 | −3 ± 52 | 25 ± 76 | 115 ± 15 | 192 ± 10 | 0.130 |
| M2-A2-G1036 | 1.1 × 1.2 | S | 202 ± 7 | | −404 ± 12 | −572 ± 15 | 209 ± 1 | 71 ± 85 | 23 ± 25 | 417 ± 35 | 98 ± 32 | 0.091 |
| M2-A3-G120 | 3.7 × 3.7 | A | 82 ± 1 | | −155 ± 6 | −240 ± 11 | 306 ± 1 | 23 ± 69 | 79 ± 93 | 268 ± 17 | 234 ± 10 | 0.169 |
| M2-A3-G1010 | 1.0 × 0.6 | S | 318 ± 12 | | −291 ± 14 | −475 ± 11 | 199 ± 1 | 41 ± 55 | 55 ± 80 | 685 ± 24 | 119 ± 12 | 0.109 |
| M2-A3-G1467 | 1.0 × 1.1 | S | 105 ± 2 | | −153 ± 7 | −265 ± 15 | 225 ± 1 | 101 ± 78 | 63 ± 97 | 94 ± 15 | 127 ± 10 | 0.092 |
| M3-G501 | 1.0 × 0.9 | S | 86 ± 2 | 38 ± 1 | −330 ± 15 | −480 ± 15 | 126 ± 1 | 81 ± 138 | 10 ± 149 | 603 ± 21 | 316 ± 11 | 0.067 |
| M3-G691 | 1.6 × 1.6 | A | 36 ± 1 | 34 ± 1 | −216 ± 8 | −230 ± 8 | 127 ± 1 | −19 ± 43 | −36 ± 44 | 114 ± 15 | −31 ± 8 | 0.044 |
| M3-G1343 | 1.2 × 1.8 | S | 52 ± 1 | 120 ± 4 | −106 ± 7 | −161 ± 7 | 8.0 ± 0.1 | 82 ± 56 | 116 ± 83 | 126 ± 41 | 120 ± 41 | 0.049 |
| M2-A3-G43 | 0.8 × 0.8 | S | 81 ± 1 | | −380 ± 7 | −500 ± 14 | 367 ± 2 | | | | | |
| M2-A3-G1683 | 0.7 × 0.8 | S | 54 ± 1 | | −335 ± 6 | −430 ± 14 | 307 ± 1 | | | | | |
| M2-A4-G1433 | 0.8 × 0.8 | S | 104 ± 6 | | −393 ± 19 | −424 ± 29 | 200 ± 1 | | | | | |
| M2-A5-G412 | 0.9 × 1.2 | S | 67 ± 2 | 22 ± 1 | −237 ± 7 | −375 ± 15 | | | | | | |

*δ notation is defined as $\delta^i A = [(^i A/^j A)_{grain}/(^i A/^j A)_{std} − 1] × 1000$, where A denotes an element, i denotes an isotope of this element, and j denotes the normalization isotope, and $(^i A/^j A)_{grain}$ and $(^i A/^j A)_{std}$ represent the corresponding isotope ratios measured in a sample and the standard, respectively.

Vanadium is less refractory than Ti (27), so this large range in $^{51}V/^{48}Ti$ is most likely caused by elemental fractionation during grain condensation. The lack of correlation between $^{51}V/^{48}Ti$ and $\delta^{49}Ti$ (Fig. 2A) and the negative correlation between $\delta^{49}Ti$ and $\delta^{30}Si$ (Fig. 2B) together imply that the grains formed after the majority of $^{49}V$ had decayed to $^{49}Ti$. Otherwise, their $\delta^{49}Ti$ values would be significantly affected by the variable amounts of live $^{49}V$ relative to $^{48}Ti$ incorporated into the grains, and no correlation would exist between $\delta^{49}Ti$ and $\delta^{30}Si$. Thus, the isotopic systematics of the X grains point to late formations of X grains, relative to the 330-day half-life of $^{49}V$, in their parent SNe and not formation within a few months after the explosions (31).

The negative correlation in Fig. 2B also implies that the $^{49}V$-$^{49}Ti$ chronometer can still provide quantitative constraints on the timing of X grain formation in SNe because the $^{49}V$ decay dominantly contributed to the $^{49}Ti$ budgets of the X grains. Large uncertainties associated with the predictions of nucleosynthetic yields from SN models, however, prevent us from accurately estimating the contributions of the Si/S zone to the $\delta^{49}Ti$ and $^{51}V/^{48}Ti$ values of individual X grains, especially given that the $^{51}V/^{48}Ti$ ratios are also affected by the fractionation of V from Ti during condensation to varying degrees. In the following discussion, a novel approach is adopted to obtain quantitative constraints on the X grain formation timing without relying on detailed model predictions.

### Subtraction of $^{49}Ti$ from the He/C zone

First, we estimate the variable amounts of material from the He/C zone that resulted in the wide range of Ti and Si isotope ratios observed in X grains. Titanium-50 is an ideal index for estimating the contribution to $^{49}Ti$ from the He/C zone because $^{50}Ti$ is not produced in the Si/S zone but is made along with $^{49}Ti$ by a neutron capture process in the He/C zone; therefore, $^{50}Ti$ in each X grain is solely from the He/C zone. In addition, Ti could be fractionated from other elements by condensation of small TiC grains before zonal mixing [for example, (37)]. The use of $^{50}Ti$ to correct for the amount of $^{49}Ti$ from the He/C zone eliminates the effect of such potential fractionation because $^{49}Ti$ and $^{50}Ti$ are isotopes of the same element. The only additional piece of information needed is the neutron capture production ratio of $^{49}Ti/^{50}Ti$ in the He/C zone, which is a function of neutron density and explosive energy. Consequently, we cannot rely on SN model predictions without first constraining the critical parameters for the neutron capture process in the progenitor SNe of the X grains.

To constrain the neutron capture environment in the He/C zone, we measured the Ti isotopic compositions of two ungrouped $^{12}C$- and $^{15}N$-enriched presolar SiC grains. Both grains had high inferred initial contents of $^{26}Al$ and $^{44}Ti$ (or $^{41}Ca$), clearly demonstrating that they formed in SNe (table S1). Isotopically distinct from X grains, both ungrouped grains show positive $\delta^{29}Si$ and $\delta^{30}Si$ values, which indicate that both grains incorporated much less $^{28}Si$-rich material from the Si/S zone than did X grains. Consequently, their Ti isotopic compositions should best represent the Ti isotopic signatures of the He/C zone that overproduces neutron-rich isotopes (Fig. 1). Both ungrouped grains lie along a 1:1 line in the plot of $\delta^{49}Ti$ versus $\delta^{50}Ti$, indicating approximately equal production of $^{49}Ti$ and $^{50}Ti$ in the He/C zones of their parent SNe. The inferred production ratio for $^{49}Ti/^{50}Ti$ of 1.04 (the solar ratio) is further supported by the Ti isotopic signature of a previously reported (38) presolar graphite grain with large $^{29}Si$ and $^{30}Si$ excesses that also originated in the SN He/C zone (39). On the basis of these observations, we adopted the production ratio of 1.04 in the He/C zones of SNe in the calculations that follow (Fig. 3) but will also discuss the uncertainty arising from this assumption below.

When corrected for neutron capture contributions based on their $^{50}Ti$ abundances, the $\delta^{49}Ti$ values of the X grains (hereafter referred to as $\delta^{49}Ti^*$; see the Supplementary Text) are more tightly correlated with their $\delta^{30}Si$ values ($R$ improves from −0.58 to −0.81), demonstrating the robustness of our approach. It is noteworthy that the adopted production ratio of 1.04 yields the highest $R^2$ value (fig. S2). The value of 1.04 represents the upper limit for the production ratio because one end of the trend after the subtraction in Fig. 3 has reached −1000‰, the minimum value for the δ notation; a higher ratio will result in $\delta^{49}Ti^*$ values below −1000‰ (that is, $^{49}Ti/^{48}Ti$ < 0) that are not physically possible. However, we cannot completely exclude the possibility of a lower $^{49}Ti/^{50}Ti$ production ratio in the He/C zone because the three ungrouped SN grains could have come from parent SNe with a different neutron capture environment in the He/C zone. Nonetheless, such a possibility can only systematically shift the negative trend in Fig. 3 to higher $\delta^{49}Ti^*$ values for the X grains (for example, fig. S3).

### Derivation of X grain formation timing

The negative correlation in Fig. 3 represents a mixing line between the Si/S zone and the He/C zone (see the Supplementary Text). Note that the trend stops at $\delta^{30}Si \approx -200‰$ because there is negligible contribution from the Si/S zone once $\delta^{30}Si$ lies above −200‰. The $\delta^{49}Ti$ value of the Si/S zone at the time of grain condensation (hereafter referred to as $\delta^{49}Ti_{Si/S}$) can, therefore, be obtained by extrapolating the negative correlation to $\delta^{30}Si = -1000‰$ (eq. S6). The corresponding $\delta^{49}Ti_{Si/S}$ value is found to be 282 ± 305‰ (95% confidence interval) by fitting a line to the grain data with the linear regression function in the Isoplot 4.15 geochronological toolkit (*40*). Supernova nucleosynthetic models predict higher Ti/Si elemental ratios in the He/C zone than in the Si/S zone (*41*), so in reality, the mixing line in Fig. 3 should be a curve with steeper slopes toward the Si/S zone. Thus, 282 ± 305‰ represents the lower limit of $\delta^{49}Ti_{Si/S}$. Instead of deriving a $\delta^{49}Ti_{Si/S}$ value with huge uncertainties based on a polynomial fit, we will adopt the lower limit of $\delta^{49}Ti_{Si/S}$ (282 ± 305‰) to more reliably constrain the X grain formation timing in the following discussion. It should be pointed out that He/C zone production ratios for $^{49}Ti/^{50}Ti$ that are smaller than we have assumed (<1.04) would shift $\delta^{49}Ti_{Si/S}$ to higher values (Fig. 2B and fig. S3), increasing the reliability of this lower limit. In this way, the lower limit for $\delta^{49}Ti_{Si/S}$ can be derived without relying on detailed SN model predictions and is therefore independent of the model uncertainties.

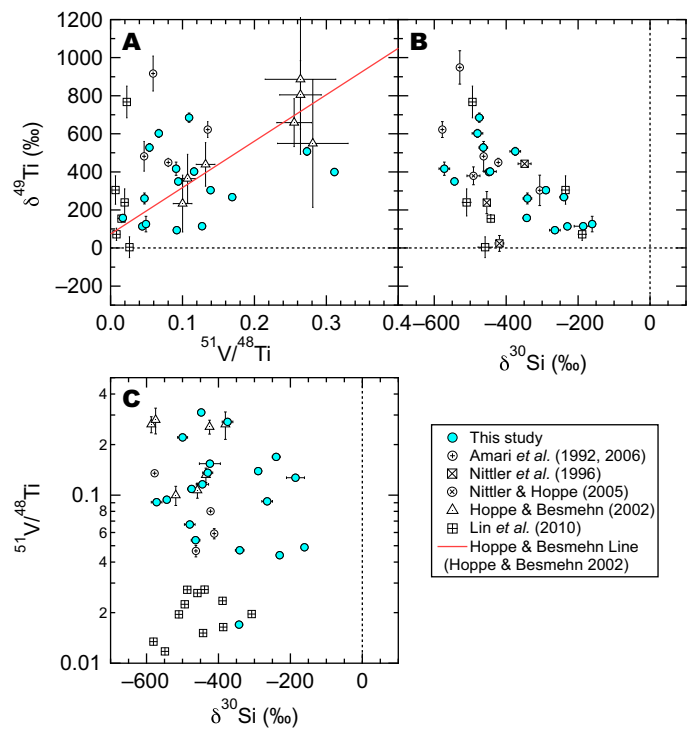

**Fig. 2. Comparison of X grain data from this study with literature data.** (**A**) $\delta^{49}Ti$ versus $^{51}V/^{48}Ti$, (**B**) $\delta^{49}Ti$ versus $\delta^{30}Si$, and (**C**) $^{51}V/^{48}Ti$ versus $\delta^{30}Si$ comparing presolar X grain data from this study (blue solid circles) with literature data [open symbols (23, 24, 31, 33, 35, 36)]. Analytical uncertainties are 1σ. Plotted in (B) are only literature data with analytical uncertainties in $\delta^{49}Ti$ (1σ error < 100‰) comparable to those in this study. See the notation of Table 1 for definition of δ values.

Recent studies (*42*, *43*) have suggested that X grains do not necessarily need to incorporate material from the inner Si/S zone to show large $^{28}Si$ excesses because a Si/C zone could form at the bottom of the He/C zone during SN explosions at high energies (for example, $5 \times 10^{51}$ erg for a $15M_\odot$ progenitor star), where the α-capture process can occur. This would produce isotopic signatures that are similar to those of the inner Si/S zone (for example, overproduction of $^{28}Si$, $^{48}Ti$, and $^{49}V$). The presence of the Si/C zone adjacent to the He/C zone is ideal for condensing SiC within a small region of an SN without the need of invoking large-scale selective mixing (Fig. 1), such as mixing the Si/S and He/C zones but not the O-rich zones in between them (*43*). However, the Ti/Si elemental ratio of the Si/C zone is predicted to be several orders of magnitude lower than that of the He/C zone (*42*). As a result, invoking the Si/C zone, instead of the Si/S zone, as the source of $^{28}Si$ predicts that $^{49}Ti$ from the He/C zone would be the dominant contributor to the $^{49}Ti$ budgets of the X grains, which is inconsistent with the grain data. Thus, we adopt SN model predictions for the Si/S zone (*41*), which is supported by the grain data, as the source of $^{28}Si$ excesses in X grains.

All the X grains show close-to-solar $\delta^{46}Ti$ and $\delta^{47}Ti$ values, indicative of solar-like metallicities for their parent SNe (*44*). Consequently, we compare our X grain results to SN model predictions (*41*) for the Si/S zones of solar metallicity progenitor stars with a wide range of initial masses in Fig. 4. It takes 2 years after the explosion to enrich the Si/S zone to a $\delta^{49}Ti$ value of −23‰, the 95% lower boundary of the $\delta^{49}Ti_{Si/S}$ lower limit. However, the typical time delay between the explosion and grain condensation in the SNe was probably even longer because the two most important potential sources of systematic uncertainties, (i) smaller $^{49}Ti/^{50}Ti$ production ratios and (ii) higher Ti/Si elemental ratios in the Si/S zone relative to the He/C zone, can shift the extracted $\delta^{49}Ti_{Si/S}$ to higher values. In support of this, we found a Ti- and V-rich subgrain with a V/Ti ratio that is three times higher than that of its host X grain (fig. S4) but observed no concomitant increase in $\delta^{49}Ti$. Given that no variation in $\delta^{49}Ti$ or $\delta^{50}Ti$ can be found between the host grain and the subgrain, the subgrain should have condensed out of the same mixed ejecta even earlier than its host grain to be captured inside [for example, (*37*)], strongly indicating the absence of live $^{49}V$ in the ejecta when the subgrain condensed. Thus, X grains must have formed at least 2 years after the SN explosions, when the majority of live $^{49}V$ had decayed to $^{49}Ti$ in the Si/S zone.

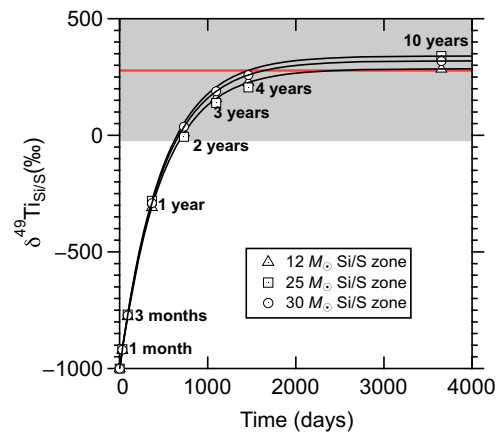

**Fig. 4. Growth curves of $\delta^{49}Ti$ in the Si/S zone resulting from $^{49}V$ decay after the SN explosions (black lines with symbols) predicted by models for solar metallicity SNe with a range of progenitor masses.** The numbers adjacent to the symbols denote the postexplosion times. The $\delta^{49}Ti_{Si/S}$ value obtained by the linear fit in Fig. 3 is shown (red line) with 95% confidence (gray band).

Finally, SN nucleosynthetic model calculations consistently predict that the maximal $\delta^{49}Ti$ value that can be reached in the Si/S zone is ~500‰, suggesting that the production ratio of $^{49}Ti/^{50}Ti$ in the He/C zone must lie above 0.5; otherwise, the $\delta^{49}Ti_{Si/S}$ values extracted from the X grains are too high to be explained (fig. S3). Because this ratio also needs to lie below unity based on the negative correlation found between the $\delta^{49}Ti*$ and $\delta^{30}Si$ values of the X grains, we can show the amount of $^{50}Ti$ produced by the neutron capture process to be about one to two times that of $^{49}Ti$ in the He/C zone.

### Implications for SN dust formation
Observations [for example, (*20*, *21*)] have suggested continuous dust formation in SNe for up to tens of years after the explosions. Our results show that SiC grains that condensed out of C-rich core-collapse SN ejecta began at least 2 years after their parent stars exploded. Although SiC is inferred to be a minor dust component according to the optical and near-infrared spectra for several SNe, with carbonaceous grains being the dominant component [for example, (*20*)], a recent SN dust condensation model predicts a delayed but concomitant formation of both C-rich phases in SNe (*32*). Our result, for the first time, provides strong support for this prediction for SiC, whereas the observation that carbonaceous grains of micrometer sizes started to appear in several SNe after a few years (20, 21) provides support for this prediction for carbonaceous grains. Like the SiC grains, the isotopic compositions of presolar SN graphite grains (the most likely form of astronomically inferred carbonaceous grains) point to outer SN zones (for example, He/C, He/N, and H envelope) as the dominant sources of the materials from which they formed (45). Presolar SN graphite and X grains, therefore, likely formed from materials from similar SN regions, providing an additional piece of evidence to support the contemporaneous formation of carbonaceous grains with SiC grains in SNe. Given the inferred delayed and concomitant formation of C-rich phases, we suggest that the observation of continuous dust formation in a few SNe over several years can be interpreted as an addition of late-forming C-rich dust to dust reservoirs in SNe. Future Ti isotopic measurements of presolar SN O-rich silicates may shed further light on this issue.

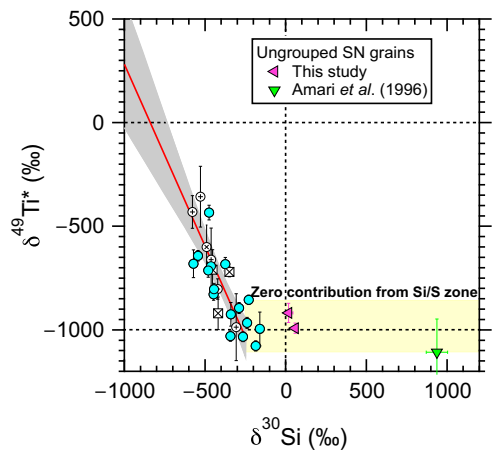

**Fig. 3. $\delta^{49}Ti*$ versus $\delta^{30}Si$ of the same set of X grains in Fig. 2 and three ungrouped SN grains.** Note that the Lin *et al.* grain data (*33*) from Fig. 2 are not plotted here because of the lack of information on $\delta^{50}Ti$ values in these grains. $\delta^{49}Ti*$ denotes $\delta^{49}Ti$ of a grain after substracting the amount of $^{49}Ti$ made by neutron capture in the He/C zone (see the Supplementary Text). The negative trend shown as a linear fit line (red solid line) with 95% confidence (gray band) results from variable contributions from the He/C zone to the $^{48}Ti$ and $^{30}Si$ budgets of the grains (eq. S5). The fact that grains with $\delta^{30}Si > \sim -200‰$ show constant $\delta^{49}Ti* = -1000‰$ (yellow region) means that the contributions from the Si/S zone are negligible in affecting the $^{48}Ti$ budgets of these grains.

## MATERIALS AND METHODS
### Sample description

The SiC grains in this study were extracted from the Murchison CM2 carbonaceous chondrite meteorite using the isolation method described in the study by Nittler and Alexander (46). About 63 g of Murchison was dissolved for presolar grain separation, and the minerals remaining after acid dissolution were first roughly separated in size based on the density difference between the CsF solution (~1.5 g/cc) and the insoluble organic matter (IOM). A small fraction (~1/63) of the IOM that efficiently traps small grains was used to extract SiC grains in this study. After destruction of the IOM, a crude-size separation based on settling times in water was conducted to reduce the number of small grains that might adhere to and contaminate the analyses of larger grains. The extracted SiC grains had a median size of 0.9 μm. A small fraction of the total SiC residue was dispersed onto high-purity gold foils attached onto aluminum stubs and subsequently pressed into the gold with an optically flat sapphire disk. Three sample mounts were prepared, referred to here as mounts #1, #2, and #3. Isotopic data of $^{13}$C- and $^{15}$N-rich presolar SiC grains on the three mounts have been reported in the studies by Liu and colleagues (47, 48).

Type X and ungrouped SiC grains investigated in this study were first nondestructively identified by their higher-than-average Mg concentrations determined by scanning electron microscope–based energy-dispersive x-ray (EDX) analyses (49) and then confirmed by isotopic analyses of C, N, and Si isotope ratios with the NanoSIMS 50L at the Carnegie Institution using standard procedures (47). A total of 82 X grains were found along with two ungrouped SiC grains with enhanced Mg concentrations according to their EDX spectra. Twenty well-isolated large X grains with a range of Si isotope ratios (fig. S1) were chosen to derive the end-member Ti isotopic compositions in the Si/S zone of their progenitor SNe. In addition, two ungrouped grains with positive $\delta^{29}$Si and $\delta^{30}$Si values were chosen to obtain Ti isotopic compositions of material from the He/C zone.

### NanoSIMS analytical procedure

For the NanoSIMS analyses, at low beam currents, the size of the Cs$^+$ primary ion beam (~1 to 3 pA) was ~100 to 150 nm for isotopic analysis of C, N, and Si as negative ions, and that of the O$^-$ primary ion beam (~10 pA) was ~300 nm for isotopic analysis of Mg-Al and Ti-V as positive ions. To minimize the problem of contamination, we used a focused ion beam (FIB) instrument to remove smaller grains adjacent to the grains of interest within 3 × 3 μm areas at the lowest beam current of 50 pA before Mg-Al and Ti-V isotopic analyses. The FIB-cleaned 3 × 3 μm areas ensured no sampling of contaminants even at increased O$^-$ beam current (up to 100 pA), which was especially important for collecting sufficient Ti$^+$ ion counts to reduce counting statistical uncertainties for the Ti isotope data. The problem of redeposited material from adjacent grains onto the grains of interest during the FIB milling was minimized by using the low-FIB beam current and also by choosing the largest X grains for this study. In addition, all the isotope data were collected in the imaging mode, which allowed selection of smaller regions of interests for isotopic ratio calculations and thus further reduced the amounts of contamination during data reduction.

The analytical procedure for isotope analyses of Mg-Al, K-Ca, Ca-Ti, and Ti-V was similar to those reported in the literature (25, 31, 33). The National Institute of Standards and Technology glass standard, SRM 610, was used as a standard for K-Ca, Ca-Ti, and Ti-V isotope analyses, whereas Burma spinel was used as a standard for Mg-Al isotope analysis. For Ti isotope analysis, synthetic TiC was used as an additional standard, and the analytical uncertainties were, for example, 13‰ and 8‰ for $\delta^{49}$Ti and $\delta^{50}$Ti, respectively, based on Ti isotope measurements of both SRM 610 and TiC standards. The 1σ errors reported in Table 1 and table S1 include both analytical and counting statistical uncertainties. The sensitivity factors of K, Ca, Ti, and V relative to Si were determined by measuring SRM 610, and the obtained values were 4.17, 9.53, 4.02, and 3.43, respectively. The sensitivity factor of Mg/Al was determined to be 1.16 in Burma spinel. All the sensitivity factors obtained in this study were consistent with the literature values within a factor of two (25, 33, 34). Note that we conducted Ti isotope analysis in the multicollection mode instead of the combined analysis mode that was used in the previous studies (24, 31, 33, 36). Although the latter allows acquisition of all the five Ti isotopes at the same time, it takes quite a while for the NanoSIMS magnet to achieve hysteresis such that it can precisely switch between two different B-field settings. Also, cycling the B-field significantly increases the time and effort taken for an analysis. Previous studies clearly show quite close-to-normal $^{46}$Ti/$^{48}$Ti and $^{47}$Ti/$^{48}$Ti ratios in X grains (24, 31, 33). In addition, in this study, we mainly aimed to obtain high-precision $^{49}$Ti/$^{48}$Ti and $^{50}$Ti/$^{48}$Ti ratios to investigate the origin of $^{49}$Ti in X grains. Thus, we avoided the combined analysis mode and measured Ca-Ti and Ti-V isotope ratios in two sets of runs in the multicollection mode: $^{28}$Si, $^{40}$Ca, $^{44}$Ca, $^{46}$Ti, $^{47}$Ti, $^{48}$Ti, and $^{51}$V in the first set of runs, and $^{28}$Si, $^{40}$Ca, $^{48}$Ti, $^{49}$Ti, $^{50}$Ti, $^{51}$V, and $^{52}$Cr in the second set. As a result, our analytical setup also allowed sufficient presputtering of potential surface contamination, for example, redeposited material during the FIB milling, in the first step and thus minimized the effect of dilution caused by surface contamination on measured $^{49}$Ti/$^{48}$Ti and $^{50}$Ti/$^{48}$Ti ratios in the second step.

## SUPPLEMENTARY MATERIALS

Supplementary material for this article is available at http://advances.sciencemag.org/cgi/content/full/4/1/eaao1054/DC1
Supplementary Text
fig. S1. Silicon three-isotope plot comparing 62 X grains found on the three gold mounts to the 20 X grains and two ungrouped grains chosen for Ti-V isotope analysis in this study.
fig. S2. $R^2$ for the correlation between the $\delta^{49}$Ti* and $\delta^{30}$Si values of X grains versus the $^{49}$Ti/$^{50}$Ti production ratios in the He/C zone showing that the smaller the production ratio, the lower the $R^2$ value.
fig. S3. The same as Fig. 3 but with $\delta^{49}$Ti* calculated by adopting a $^{49}$Ti/$^{50}$Ti production ratio of 0.50 instead of 1.04.
fig. S4. $\delta^{49}$Ti (upper panel) and $^{51}$V/$^{48}$Ti (lower panel) versus NanoSIMS analysis cycle showing the inclusion of a Ti- and V-rich subgrain (gray region) within X grain M1-A8-G138, which had enhanced V/Ti ratios but no concomitant increase in $\delta^{49}$Ti within analytical uncertainties (~100‰), indicating incorporation of a negligible amount of live $^{49}$V, that is, formation after the decay of the majority of live $^{49}$V in the Si/S zone.

table S1. Carbon, N, Si, Al, K-Ca, and Ca-Ti isotopic compositions of two ungrouped SiC grains and one ungrouped graphite, KE3d-9 (38), separated from Murchison.

**Acknowledgments**
**Funding:** This work was supported by NASA (grants NNX10AI63G and NNX17AE28G to L.R.N.). **Author contributions:** N.L. designed the research; C.M.O'D.A. prepared the samples; N.L. and J.W. performed the research; N.L. and L.R.N. interpreted the data; and N.L., L.R.N., and C.M.O'D.A. wrote the paper. **Competing interests:** The authors declare that they have no competing interests. **Data and materials availability:** All data needed to evaluate the conclusions in the paper are present in the paper and/or the Supplementary Materials. Additional data are available from authors upon request.

Submitted 14 June 2017
Accepted 13 December 2017
Published 17 January 2018
10.1126/sciadv.aao1054

**Citation:** N. Liu, L. R. Nittler, C. M. O'D. Alexander, J. Wang, Late formation of silicon carbide in type II supernovae. *Sci. Adv.* **4**, eaao1054 (2018).